\documentclass[prl,aps,twocolumn]{revtex4}
\usepackage{graphicx}
\usepackage{amsmath}
\usepackage{bm}

\newcommand{\be}{\begin{equation}}
\newcommand{\ee}{\end{equation}}
\newcommand{\bea}{\begin{eqnarray}}
\newcommand{\eea}{\end{eqnarray}}
\newcommand{\bei}{\begin{itemize}}
\newcommand{\eei}{\end{itemize}}

\begin{document}

\title{Observable Bulk Signatures of Non-Abelian Quantum Hall States}

\author{N. R. Cooper$^{(1)}$ and Ady Stern$^{(2)}$}
\affiliation{(1) Theory of Condensed Matter Group, University of Cambridge, Cavendish Laboratory, J.~J.~Thomson Ave., Cambridge CB3~0HE, UK.}
\affiliation{(2) Department of Condensed Matter Physics, Weizmann
Institute of Science, Rehovot 76100, Israel.}

\begin{abstract}

  We show that non-abelian quantum Hall states can be identified by
  experimental measurements of the temperature-dependence of either the
  electrochemical potential or the orbital magnetization. The predicted
  signals of non-abelian statistics are within experimental resolution, and
  can be clearly distinguished from other contributions under realistic
  circumstances. The proposed measurement technique also has the potential to
  resolve spin-ordering transitions in low density electronic systems in the
  Wigner crystal and strongly-interacting Luttinger liquid regimes.

\end{abstract}
\date{20 December, 2008}
\pacs{}
\maketitle

Recent interest in many-body systems whose quasi-particles satisfy non-abelian
quantum statistics has been stimulated by strong theoretical arguments that
such states may be realized in certain fractional quantum Hall effect states,
and by the possible use of non-abelian quasi-particles for topological quantum
computation\cite{nayakrmpsternannals}. In view of this interest, it is widely accepted
that an experimental test of non-abelian quasi-particle statistics is highly
desirable.  

The fractional quantum Hall states within the Landau filling range of
$2<\nu<3$ are the most accessible candidate states for such a test. Of these,
the most prominent is the $\nu=5/2$ state\cite{PhysRevLett.59.1776}, which
is believed likely to be described\cite{morfpfaffian,PhysRevLett.84.4685} by the Moore-Read state\cite{MooreR91} or a
closely related state\cite{levinantipfaffian,lee:236807}.
Experiments have shown the quasi-particles of the $\nu=5/2$ state to have
fractional charge $e/4$\cite{heiblumeover4marcuseover4}, consistent with these theoretical
expectations. However, existing measurements do not discriminate between these
non-abelian phases and other paired phases with simple abelian statistics.

Non-abelian quantum states are characterized by two properties:
first, in the presence of well separated quasi-particles, the
ground state is degenerate, and the degeneracy is exponential in
the number of quasi-particles. And second, an adiabatic  braiding
of the positions of quasi-particles results in a unitary
transformation that takes the system from one ground state to another,
and that depends only (up to an overall phase) on
the topology of the braiding. Previously proposed experimental
tests have mostly focused on the use of interferometers to probe
the effect of braiding\cite{sarmafn,sternh,bonderson:016803,feldman:085333}. Largely, they make use of quasi-particles
flowing along edges and encircling quasi-particles that are
localized in the bulk. Remarkably, an interferometer has been proposed for utilization as a topological qubit\cite{sarmafn}.

In this Letter, we propose experimental tests of the exponential degeneracy of
the ground state through the effect it has on the low temperature entropy. While in principle this entropy may be probed directly by a
measurement of the system's heat capacity, in practice the heat capacity of
the two-dimensional electron gas (2DEG) is negligible relative to that of the
three dimensional phononic system. The tests we propose measure only the
electronic contribution. When compared to previously proposed interferometric tests, they enjoy the advantage of not relying on any assumed structure of the edge, and hence not being sensitive to edge reconstructions, non-equilibrium edge distributions etc. 
While not probing braiding, they do probe the existence of non-abelian
quasi-particles. Therefore, they can distinguish a phase of the Moore-Read
type from an abelian phase, or a phase where the $e/4$ excitations pair into
abelian $e/2$ quasi-particles \cite{pairedrefs}.
A previously suggested test of the ground state degeneracy has focused on its effect on the thermopower\cite{kunyang}. Being a transport property, however, one may worry whether it survives the effect of quasi-particle localization.

For the brevity of our presentation, we focus on the
signatures of the proposed Moore-Read state at $\nu=5/2$. The results for the particle-hole conjugate state (the ``anti-Pfaffian'') are the same. The generalization to
other non-abelian quantum Hall states is described at the end.
For the Moore-Read state with $N_{\rm qp}$ quasi-particles at well-separated
fixed locations, there are $2^{N_{\rm qp}/2-1}$ ground states, that may be
described by different occupations of the Majorana modes\cite{ReadG00}. Tunnel
coupling of the Majorana modes splits the degeneracy, and introduces a small
energy scale $J$ that is exponentially suppressed in the ratio of the
quasi-particle spacing to the correlation length of the groundstate (presumably
of the order of the magnetic length). We shall assume throughout that $J\ll
k_{\rm B} T$. We note that although $J$ is small, it is important that it is
  non-zero, such that on the timescale of the measurements, the Majorana modes
  can come to thermal equilibrium. The Majorana modes therefore result in a
temperature-independent contribution to the entropy per unit area, of size \be
s_{\rm MR} = k_{\rm B} \; n_{\rm qp}\; \frac{1}{2} \ln 2 
\label{eq:nonabent}
\ee
where $n_{\rm qp}$ is the density of quasi-particles.
Since the quasi-particles have charge $e/4$, this density is
\be n_{\rm qp} = 4 \left|n - \frac{5n_0}{2}\right| \label{eq:nqp}
\ee
where
$n$
is the areal density of electrons and  $n_0
\equiv eB/h$ is the Landau level degeneracy for  
magnetic field $B$ perpendicular to the 2DEG.

The thermodynamic signature of the non-abelian quantum Hall state is therefore
the
change in the entropy density (\ref{eq:nonabent}) with changing number density
of quasi-particles (\ref{eq:nqp}), which may be effected by varying either the
density of electrons $n$
or the magnetic field $B$.
Central to our proposals is the observation that this change in entropy density
has direct consequences for the electrochemical potential $\mu$ and the
magnetization density $m$, as may be deduced by employing
the Maxwell relations
\bea
\label{eq:maxwellm}
\left(\frac{\partial m}{\partial T}\right)_B & = &
\left(\frac{\partial s}{\partial B}\right)_T  \\
\left(\frac{\partial \mu}{\partial T}\right)_n & = & - \left(\frac{\partial
    s}{\partial n}\right)_T \,.
\label{eq:maxwellmu}
\eea
If the $\nu=5/2$ quantum Hall state is in the phase described by the
Moore-Read state, then, using the Maxwell relations
(\ref{eq:maxwellm},\ref{eq:maxwellmu}) together with Eqns.(\ref{eq:nonabent},\ref{eq:nqp}),
we find that the non-abelian statistics of the quasi-particles lead to the
contributions: \bea \left(\frac{\partial m_{\rm MR}}{\partial
      T}\right)_{n,B} \!\!\! & = & \frac{ds_{\rm MR}}{d n_{\rm qp}}
\left(\frac{\partial n_{\rm qp}}{\partial B}\right)_{n,T} \!\!\! = \mp\frac{k_{\rm B}
  e}{h} 5 \ln 2
\label{eq:dmdt}
\\
\left(\frac{\partial \mu_{\rm MR}}{\partial T}\right)_{n,B} \!\!\! & = & -\frac{ds_{\rm
      MR}}{d n_{\rm qp}} \left(\frac{\partial
      n_{\rm qp}}{\partial n}\right)_{B,T} \!\!\! =
\mp k_{\rm B}2
\ln 2
\label{eq:dmudt}
\eea where the signs on the right hand side correspond to the cases $\nu
\stackrel{>}{<} 5/2$ of quasi-electrons or quasi-holes. The discontinuity at
$\nu=5/2$ is characteristic of a clean system. In the presence of disorder,
the ground state at $\nu=5/2$ would include pinned quasi-electrons and
quasi-holes (of equal average density), that would smooth the transition
through zero at $\nu=5/2$.

{We propose that the thermodynamic signal of non-abelian quasi-particles be
  sought in measurements of the magnetization density or of the
  electrochemical potential at fixed density}. Measurements of these quantities do not
suffer from background contributions from the three dimensional substrate
around the quantum Hall system. We now show that these quantities can be
measured to a degree of accuracy that should allow resolution of the
non-abelian contribution to the entropy (\ref{eq:nonabent}). Following that, we estimate other
sources of entropy that may mask the signal we are after. We show that these sources are likely to be smaller in magnitude, and different in temperature dependence, than the contribution of the internal degrees of freedom of the non-abelian quasi-particles.

The predicted signal of
the non-abelian quasi-particles of the $\nu=5/2$ state on the electrochemical potential is a
temperature-dependent shift $\Delta\mu_{\rm MR}
\simeq 1.4 \Delta(k_{\rm B} T)$, Eqn.(\ref{eq:dmudt}), when the measurement is taken with a fixed density. There exist various techniques for measuring
the variations of the chemical potential of a
2DEG\cite{eisensteincompress,IlaniYMS00,reznikov,ilani-2004-427}, which have allowed
studies of both the average\cite{eisensteincompress} and local
compressibilities\cite{ilani-2004-427} in the fractional quantum Hall
regime.
The electrochemical potential $\mu$ is defined by the voltage drop between the
electron gas and ground. A small change in temperature $\Delta T$ at
fixed electrochemical potential ($\Delta\mu=0$) will, in general, lead
to a small change in the electron density, $\Delta n$. This change $\Delta n$ could be detected
by a measurement of the charge that flows onto the 2DEG\cite{reznikov}, or by the change in the electrostatic potential around the 2DEG, for example by a
Single Electron Transistor (SET) implanted on the top of the
device\cite{ilani-2004-427}. Now, by making a small adjustment $\Delta\mu$ to the voltage
between the ground and the 2DEG, one can use these measurements to eliminate the
change in electron
density ($\Delta n =0$). The ratio $\Delta \mu/\Delta T$ then provides a
measure of $(\partial\mu/\partial T)_n$, Eqn.({\ref{eq:dmudt}). The
  electrostatic potential can be measured with a sensitivity of order
  $1\mbox{mK}$ by an SET\cite{ilanicomm}, allowing changes in $\mu$ to be
  measured with milli-Kelvin accuracy. 
Thus, the predicted signal $\Delta\mu_{\rm MR} \simeq 1.4
  \Delta(k_{\rm B} T)$, Eqn.(\ref{eq:dmudt}), could be measured even for
  temperature variations on the scale of a few milli-Kelvin, well below the
  bulk gap of the $\nu=5/2$ state\cite{choi:081301,dean:146803}. On this temperature scale a very small
  value of $\rho_{xx}$ is experimentally observed at and near $\nu=5/2$,
  indicating that quasi-particles are immobile.

The predicted signal of non-abelian statistics of the $\nu=5/2$ quasi-particles
in the magnetization density is a temperature-dependent change of size $\Delta
m_{\rm MR} \simeq 12 \mbox{pA} \times (T/\mbox{mK})$, Eqn.(\ref{eq:dmdt}). The
magnetization of 2DEGs can be sensitively measured by torque
magnetometry methods\cite{wiegersprl,PhysRevLett.82.819,schwarz,harris}. By
the use of micromechanical cantilevers\cite{schwarz,harris} a sensitivity of
order $\Delta m \simeq 400 \mbox{pA}$ has been reported\cite{ruhe:235326}, while achievable force sensitivies of $\sim 10^{-17} \mbox{N}$ can allow $\Delta m \simeq 0.3\mbox{pA}$ 
\cite{harriscomm}. Thus, the predicted signal
should be observable for a temperature change as small as a few
milli-Kelvin, again well below the energy gap of the
$\nu=5/2$ FQH state.

Some further comments are in order regarding the use of magnetization to
measure the entropic contributions.

The magnetization of an electron gas has both orbital and spin contributions.
For electrons in GaAs quantum wells
torque magnetometry is sensitive only to the orbital contribution
to the magnetization. 
This is convenient for our
purposes, as it is the orbital magnetization which shows
the signature of the non-abelian nature of the quasi-particles.

For a two-dimensional sample, the magnetization density $m$ is the
total current flowing around the sample edge. It is perhaps surprising that
this edge current can be affected by the Majorana degrees of freedom of the
bulk quasi-particles, which are electrically neutral and located far from the
edge itself.
Some intuition to this effect may be obtained by regarding the bulk and the
edge as two systems at mutual thermal equilibrium. As the temperature is, say,
increased and the filling factor is slightly above $\nu=5/2$, the
electrochemical potential of the bulk decreases as a result of the
quasi-particle entropy (\ref{eq:nonabent}). (The electrochemical potential is
the change in free energy associated with adding a single electron, i.e., four
quasi-electrons, at fixed temperature.) To equalize its electrochemical
potential to that of the bulk, the edge then needs to lose electrons. Since the electrons in the edge region experience classical skipping orbits that give a paramagnetic contribution to the magnetization, the
loss of electrons from the edge reduces the magnetization density.

While our analysis above shows that the entropy (\ref{eq:nonabent}) of the
quasi-particles of the $\nu=5/2$ state has detectable effects within available
possible measurements, its actual observation  depends on its relative
magnitude to other sources of entropy. Since the measurements we propose are
sensitive only to the electronic entropy, and since the bulk quantum Hall
state is gapped, the main source of entropy is the positional degrees of
freedom of the quasi-particles. We now turn to estimate this contribution.

We consider first the positional entropy of  non-interacting
quasi-particles.
Exact formulas have been derived for a model Hamiltonian for which the
Moore-Read state and its quasi-holes are exact zero energy states on a
finite spherical geometry\cite{ReadR96}. (Similar exact formulas have
been derived also for the more complicated Read-Rezayi
states\cite{Ardonne,Read06}.)  These formulas provide the total
degeneracy of the ground state -- including both internal and
positional degrees of freedom of the quasi-particles -- and hence, by
taking the thermodynamic limit, the total entropy density of the
system.  By this calculation, the unique contribution of the
non-abelian nature of the quasi-particles is masked by the positional
entropy, with $s \sim k_{\rm B} n_{\rm qp} \ln(n/{n_{\rm qp}})$.

These formulas cannot hold when the QHE is observed, as they
neglect the interaction of quasi-particles with one another and with a
disorder potential, which are essential for quasiparticle
localization.
These interactions act to constrain the motion of the quasi-particles,
and as a consequence introduce a temperature dependence to the
positional entropy which will make it possible to distinguish it from
the entropy we are interested in. The exponentially small $J$ led to
the $T$-independent expressions (\ref{eq:dmdt}) and
(\ref{eq:dmudt}). In contrast, no such exponentially small band width
exists for the positional degrees of freedom. 

In the low density limit, the quasi-particles will be pinned by
impurities, with an approximately uniform density of states per unit
energy for the range extending from zero energy to the energy gap
$E_g$ (the energy corresponding to a creation of a
quasi-electron/quasi-hole pair). The positional entropy would then be
proportional to $T/E_g$. (We assume hard-core quasi-particle
repulsion, to prevent quasi-particles accumulating in a single
potential well.)

When the density is higher and the system is clean enough that the
separation between quasi-particles is smaller than the correlation
length of the disorder potential, crystallites of quasi-particles will
be formed, with phononic excitation spectrum.  As in the case of
disorder, the entropy will again be temperature dependent.
For Coulomb-interacting quasi-particles the crystallites will form
triangular Wigner lattices, with lattice constant $a$ set by $n_{\rm
qp} \equiv {2}/{\sqrt{3} a^2}$. In a magnetic field it is
characterized by the Debye energy $E_D \sim
\frac{e^2}{\epsilon\epsilon_0 a}\frac{n_{qp}}{n_0}$. For $k_B T \ll
E_D$, the positional entropy arises from the thermal excitation of the
long-wavelength phonons of the Wigner crystal. The magnetophonon mode
has dispersion $E_k = E_D (ka)^{3/2}$\cite{chaplik,fisher}.  Thermal
excitation of these phonons leads, in the low temperature limit, to a
positional contribution to the entropy of
\be s_{\rm WC}  \stackrel{T\to 0}{\simeq}  \frac{7 k_{\rm
B}}{9\pi a^2} \Gamma\left({4}/{3}\right)\zeta\left({7}/{3}\right)
\left(\frac{k_{\rm B} T}{E_D}\right)^{4/3}  \sim  k_{\rm B} n_{\rm qp} \left(\frac{k_{\rm B} T}{E_D}\right)^{4/3} \,.
\label{eq:swc}
\ee
The entropy associated with the positional degrees of freedom is given
by Eqn.(\ref{eq:swc}) for $k_{\rm B}T\ll \mbox{min}(E_D,E_m)$, with
$E_m$ being the melting temperature of the lattice.
Even at higher temperature, when the solid melts, the entropy is
expected to vary strongly with temperature, up to temperatures of the
scale of the mean interaction energy between quasi-particles.
Provided the temperature is below this scale but large compared
to $J/k_{\rm B}$, the non-abelian entropy (\ref{eq:nonabent}) will be
the only temperature independent entropy.

Thus, for samples in which the positional entropy of the quasi-particles arises
largely from the phonons of a Wigner lattice, it should be possible to observe
clear evidence of the non-abelian statistics in the measurements we propose.
For large deviations of filling factor from $\nu=5/2$ the Wigner crystal of quasi-particles competes with other phases, including a ``stripe'' phase of electrons\cite{MoessnerC96KoulakovFS96}. A first order transition into such a phase will lead to singularities in the derivatives of the entropy density $s$. In the vicinity of this transition, one expects a first order transition as a function of temperature, owing to the differing entropies of the Wigner crystal of quasi-particles and the competing stripe phase. In the measurement schemes that we propose this phase transition will appear as singularities in $(\partial m/\partial T)_B$ and $(\partial\mu/\partial T)_n$.

The extension of this calculation from $\nu=5/2$ to the other states of the $\nu=2+\frac{k}{k+2}$ Read-Rezayi series\cite{ReadR99} is straightforward. Eqn.(\ref{eq:nqp}) reads $n_{\rm qp} = (k+2) \left|n - \nu n_0\right|$. For a large $N_{\rm qp}$ the degeneracy of the ground state approaches $D^{N_{\rm qp}}$, with the quantum dimension $D=2\cos{\pi/(k+2)}$\cite{freedman2004}. Consequently, $\left(\partial m_{\rm RR}/\partial T\right)_B=\mp (k_{\rm B}e/h) \nu(k+2)\ln{D}$, and $\left(\partial\mu_{\rm RR}/\partial T\right)_n=\mp  k_{\rm B} (k+2)\ln{D}$.

Before concluding, we note that very similar physics arises under very
different circumstances. Consider the $d$-dimensional electron gas in zero (or
weak) magnetic field, in the regime of low particle density (large $r_s$)
where the electron gas is expected to behave as a classical Wigner crystal.
Deep inside this Wigner crystal regime, the electrons are localized close to
the sites of the classical groundstate and exchange interactions are
suppressed. There is then a wide separation of energy scales, between the characteristic interaction scale set by
the typical Coulomb energy $E_C \sim e^2 n^{1/d}/(\epsilon \epsilon_0)$ and the
energy scale set by spin exchange interactions $J'$, which is exponentially small. For $J'\ll k_{\rm B}T \ll
E_C$: (i) the temperature is sufficiently high that magnetic order is
destroyed, so there is a entropy density arising from spin of
\be
s_{\rm spin}(k_{\rm B}T \gg J') = k_{\rm B} n \ln 2 \; ,
\label{eq:spinent}
\ee
(ii) the positional entropy of the particles is strongly suppressed, being
described by phonons of the Wigner lattice and adopting a power-law
temperature dependence for $k_BT$ much less than the Debye energy, which in this case is $E_D \sim E_C/\sqrt{r_s}$. For the 2D Wigner crystal, the long-wavelength phonons (2D
plasmon) have the dispersion $E_k \sim k^{1/2}$, leading to $s_{\rm WC} \sim
n k_{\rm B} (k_{\rm B}T/E_D)^4$. In one dimension, this regime $J'\ll k_{\rm B}T
\ll E_C$ is referred to as the ``spin-incoherent Luttinger liquid''
regime\cite{matveev,fiete}, for which one finds $s_{\rm WC} \sim n k_{\rm
  B} (k_{\rm B} T/E_D)$ at low temperatures. Thus, for $k_{\rm B}T \ll E_D$ the spin entropy
(\ref{eq:spinent}) is the dominant contribution. Using measurements of the
temperature-dependence of the electrochemical potential (\ref{eq:maxwellmu})
as described above, one can expect to observe experimentally the spin
contribution to the entropy and transitions associated with changing magnetic
order under variations of $J'/(k_{\rm B}T)$, via
\be
\left(\frac{\partial \mu}{\partial T}\right)_n  =  - \left(\frac{\partial
    s}{\partial n}\right)_T  \simeq  -\left(\frac{\partial
    s_{\rm spin}}{\partial n}\right)_T
\ee
for $k_{\rm B}T \ll E_D$.
Studies of this type could provide important
thermodynamical information on the nature of spin-ordering in quasi-1D and
quasi-2D electron states at low densities.
Such measurements are related to the technique described in Ref.\cite{reznikov}. By the use of measurements of electrostatic potential by an SET, these techniques allow access to small structures (e.g. quasi 1D wires).

In summary, we have shown that experimental evidence for non-abelian
quantum Hall states can be obtained from measurements of the magnetization
density and electrochemical potential. 
These quantities allow the determination of the
entropy associated with the non-abelian degrees of freedom of the
quasi-particles. The predicted signal of non-abelian statistics is within
experimental resolution, and can be clearly resolved from other entropic
contributions at sufficiently low temperature. 
Our suggested measurement
technique also has the potential to resolve the thermodynamic signatures of
spin-ordering transitions in low density electron gases in the Wigner crystal
and strongly-interacting Luttinger liquid regimes.

\vskip0.5cm

\acknowledgments{We are grateful to J.P. Eisenstein, B.I. Halperin, Jack Harris, Shahal Ilani, Nick Read, Amir Yacoby and in particular Kun Yang for helpful
  comments. This work was partially supported by EPSRC Grant
  No. EP/F032773/1, by the US-Israel bi-National Science Foundation and by Microsoft's Station Q.}


\end{document}